\documentclass[preprint,superscriptaddress,prd,eqsecnum]{revtex4}
\usepackage{amssymb,amsmath,graphicx,amscd,simplewick,xcolor}
\usepackage[export]{adjustbox}
\usepackage{float}
\begin{document}

\title{Vacuum Decay Induced by Quantum Fluctuations}

\author{Haiyun Huang}
\email{jasmine.haiyun@gmail.com}
\affiliation{Institute of Cosmology, Department of Physics and Astronomy \\
Tufts University, Medford, Massachusetts 02155, USA}

\author{L. H. Ford}
\email{ford@cosmos.phy.tufts.edu}
\affiliation{Institute of Cosmology, Department of Physics and Astronomy \\
Tufts University, Medford, Massachusetts 02155, USA}

\begin{abstract}
We treat the effects of quantum field fluctuations on the decay of a meta-stable state of a self-coupled
scalar field. We consider two varieties of field fluctuations and their potential effects in a semiclassical description.
The first are the fluctuations of the time derivative of a free massive scalar
field operator, which has been averaged over finite regions of space and time.  These fluctuations obey a Gaussian
probability distribution. A sufficiently large fluctuation is assumed to produce an effect analogous to a classical
initial field velocity, which can cause a finite region
to fly over the barrier separating the meta-stable state from the stable vacuum state. Here we find a contribution
to the decay rate which can be comparable to the decay rate by quantum tunneling, as computed in an instanton
approximation. This result is consistent with those of other authors. We next consider the effects of the
fluctuations of operators which are quadratic in the time derivative of the free scalar field. The quadratic operator
is also averaged over finite regions of space and of time. Now the probability distribution for the averaged operator
falls more slowly than an exponential function, allowing for the possibility of very large fluctuations. We find a
contribution to the decay rate which, under certain conditions, may be larger than those coming from either quantum tunneling or
linear field fluctuations.
\end{abstract}

\maketitle
\baselineskip=14pt	

\section{Introduction}

Quantum tunneling is a well known effect in quantum mechanics. For example, a quantum particle in
a local potential minimum has a nonzero probability to tunnel through a potential maximum to reach another
potential minimum with lower energy. In many cases, the tunneling amplitude may be accurately calculated
using the WKB approximation. However, there can be quantum field theoretic corrections to the tunneling rate
calculated in single particle quantum mechanics. If the particle has an electric charge, it will respond to vacuum
fluctuations of the electric field, which will result in a small increase in the tunneling rate~\cite{ FZ99,HF15}.
This increase arises from a one-loop correction to the tree level scattering amplitude, which here could be the
WKB tunneling amplitude. If the particle is an electron, then this correction is described by the one-loop vertex
diagram in quantum electrodynamics. However, as was discussed in Ref.~\cite{ HF15}, the increase in tunneling
rate may be reasonably estimated from a simple semiclassical argument. The particle is subjected to vacuum
electric field fluctuations even if no real photons are present. These fluctuations exert a force on the particle
which can either push the particle toward the barrier, enhancing the tunneling probability, or away from the barrier,
suppressing tunneling. However, the average effect is a small enhancement of the tunneling rate, in  agreement
with the one-loop perturbation theory calculation.

 Vacuum radiation pressure fluctuations can also enhance the  transition rate, and were studied in Ref.~\cite{HF16}.
 Here the enhancement is potentially larger, and can possibly dominate over the effect predicted by
 the WKB approximation. This type of large vacuum fluctuation will be discussed in more detail below in Sect.~\ref{sec:quad}.
 Vacuum radiation pressure fluctuations on Rydberg atoms were recently discussed in Ref.~\cite{F21}, where it
 was argued that these fluctuations might produce observable effects.

 The topic of the present paper will be role of quantum field fluctuations in the decay of  a false vacuum state in
 field theory. Consider a real scalar field, $\phi({\bf x},t)$, with self-coupling described by a potential, $U(\phi)$,
 which  has at least two local minima. If we quantize small perturbations around the global minimum, the ground state
 of the resulting field theory is  called the true vacuum, whereas if we select a minimum with higher energy, the
 corresponding state is called a false vacuum, and is potentially unstable against decay into the true vacuum.
 A  Euclidean space formalism which describes this decay by quantum tunneling was developed by Coleman~\cite{C77},
 and will be reviewed in Sect.~\ref{sec:instanton}.

 The outline of this paper is as follows: Section~\ref{sec:instanton} will review false vacuum decay by quantum tunneling,
 as described in the instanton approximation.  Section~\ref{sec:class} will discuss the classical dynamics of a self-coupled
 scalar field with two local potential minima, and illustrate how suitable initial conditions on the time derivative,
 $\dot{\phi}$, of the  classical field can cause a  finite region to fly over the potential maximum separating these minima.
 In Sect.~\ref{sec:phidot}, we consider the effects of the vacuum fluctuations of the linear field operator, $\dot{\phi}({\bf x},t)$,
 averaged over a finite spacetime region, and argue that this effect can be of the same order as the instanton contribution
 to the decay rate. A similar conclusion was reached some time ago by Linde~\cite{Linde92}, who studied the effects of
 ${\phi}$ fluctuations in de Sitter spacetime. The effects of linear field fluctuations were also studied by Calzetta, Verdaguer,
 and coworkers~\cite{CV99,CRV02,CRV01,ACRV03,CV06} in several models, and agree with Linde's conclusion.
 After an early version of our results was first  presented~\cite{seminar18}, we
 become aware of the recent work of several authors~\cite{BJPPW19,HY19,BDV19,Wang19}, who treat either  ${\phi}$  or $\dot{\phi}$
 fluctuations in either Minkowski or de Sitter spacetime, and also conclude that the contribution to the decay rate is of the same order
 as the instanton contribution. However, these authors disagree as to whether linear quantum field fluctuations and
 instanton methods are different formalisms for describing the same physical process, or whether they describe physically
 distinct processes. This is a question to which we will return later in this paper. Section~\ref{sec:quad} will discuss the
 fluctuations of a spacetime average of the quadratic operator ${\dot\phi^2}$. We first review results from Refs.~\cite{FFR12,FF15,SFF18,FF19,WFS21}
 to the effect that the probability distribution for such an operator can fall more slowly than an exponential function.  We then
discuss whether  the effects of ${\dot\phi^2}$ fluctuations on the
 decay rate of the false vacuum can be larger than either quantum tunneling, as described by an instanton, or
 the effects of linear field fluctuations. Our results are summarized and discussed in Sect.~\ref{sec:final}

 Units in which $\hbar = c =1$ will be used throughout this paper.

\section{Instanton Methods }
\label{sec:instanton}

The instanton method approximates a path integral in euclidean space as being dominated by one or more solutions
of locally minimum euclidean action, the instantons. This leads to an expression for a transition amplitude in the form
of a sum of terms of the form $\exp(-S)$, where $S$ is the  euclidean action of an instanton. This method is analogous
to the use of the saddle point or stationary phase approximations for the evaluation of ordinary integrals, and is reviewed
by Coleman in Ref.~\cite{C85}.

\subsection{Quantum Mechanics and the Schwinger Effect}
\label{sec:QM}

Instanton methods may be used to compute barrier tunneling rates in single particle quantum mechanics. The results are similar to
those from the WKB approximation. The instanton and WKB methods have been compared by several
authors~\cite{Dunne2000,M-K01,Benderskii08}, who find that the two methods are not identical, but often give answers
which agree  to reasonable accuracy.

Instanton methods may also be applied to the Schwinger effect~\cite{Schwinger51}, the creation of pairs of charged particles and antiparicles
from the vacuum by a constant electric field. This was done by Garriga~\cite{Garriga94a} both in a $1+1$ and in a $3+1$ dimensional models.
 The $3+1$ dimensional case was treated in more detail by Kim and Page~\cite{KP02}. The result for the creation rate obtained by  instanton
approaches agree with those found by Bogolubov coefficient methods in  $1+1$~\cite{Garriga94b} and in $3+1$ dimensions~\cite{Grib94},
as well as  with Schwinger's original approach using an effective action.  Thus the instanton approach seems to give a reliable description
of the Schwinger effect.

\subsection{Instantons in Quantum Field Theory and False Vacuum Decay}
\label{sec:QFT}

\begin{figure}[htbp]
\includegraphics[scale=0.6]{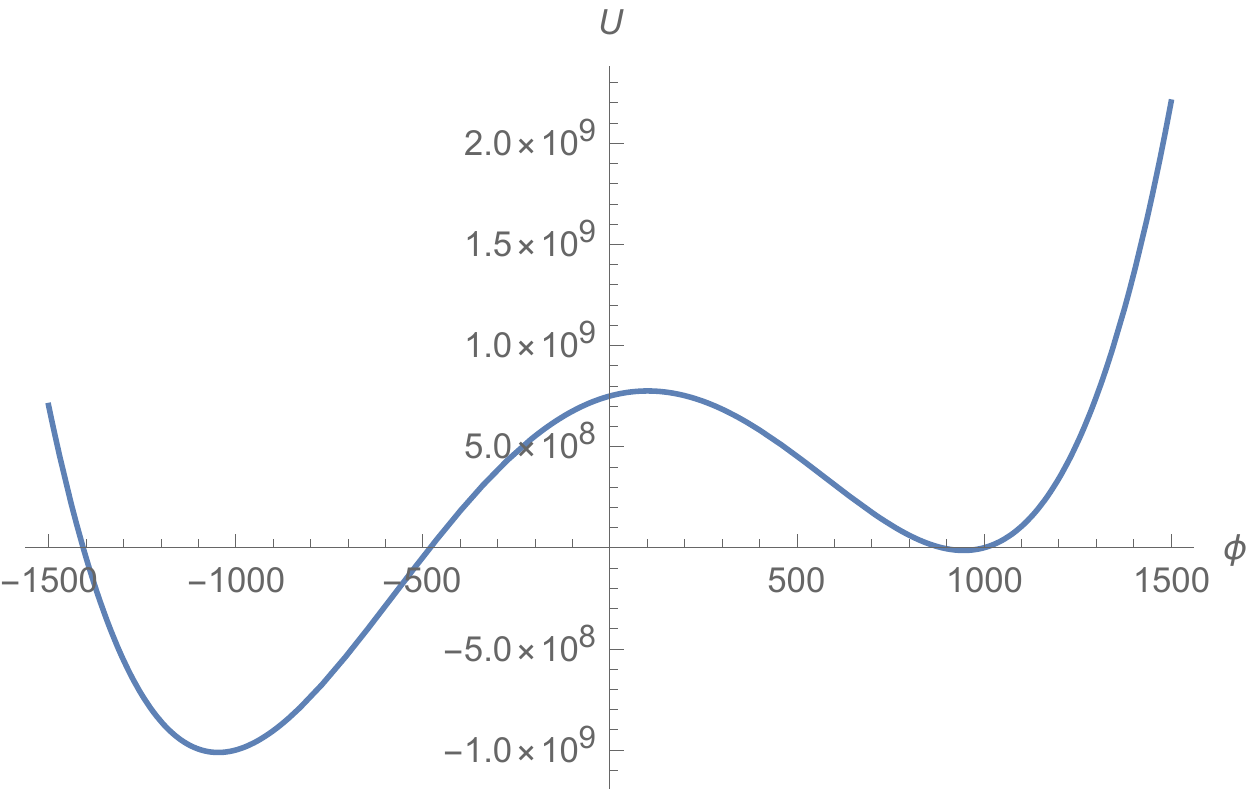}
\caption{The potential $U(\phi)$ given in Eq.~\eqref{eq:potential}  is plotted with the choice of parameters in Eq.~\eqref{eq:parameters}.}
\label{fig:potential}
\end{figure}

In this section, we summarize the instanton method used by Coleman~\cite{C77} to estimate the rate of false vacuum decay.
Consider a real scalar field with the Lagrangian density
\begin{align}
\mathcal L=\frac{1}{2}\partial_\mu \phi\partial^\mu\phi+U(\phi),
\label{eq:Lagrange}
\end{align}
where $U(\phi)$ is a ``double well'' potential with two minima. The associated equation for $\phi({\bf x},t)$ is
\begin{equation}
\Box \phi - U'(\phi) = 0\,,
\label{eq:eom}
\end{equation}
where $\Box$ denotes the d'Alembertian operator in Lorentzian space [where we use metric signature $(-,+,+,+)$],
 and the four-dimensional Laplacian in Euclidean space.
A specific choice for  $U(\phi)$  is
\begin{align}
U(\phi)=\frac{\lambda}{8}(\phi^2-a^2)^2+\frac{\epsilon\lambda a^3}{2}(\phi-a)\,,
\label{eq:potential}
\end{align}
where $\lambda$, $a$, and $\epsilon$ are positive real constants. This form is illustrated in Fig.~\ref{fig:potential}  for the case
\begin{equation}
\lambda=0.01\,, \quad a=1000\,,  \quad  \epsilon=0.1\,.
\label{eq:parameters}
\end{equation}

The potential has a local minimum at $\phi=\phi_+$, the false vacuum, and a global minimum at $\phi=\phi_-$, the true vacuum. These minima are
separated by a local maximum at $\phi=\phi_m$. The potential difference between the false vacuum and the local maximum is
$\Delta U = U(\phi_m) - U(\phi_+).$
 For the choice of parameters given in  Eq.~\eqref{eq:parameters} and illustrated in Fig.~\ref{fig:potential},
$\phi_- \approx -1046.7$, $\phi_m \approx 101.0$, and $\phi_+\approx 945.6$. Thus
\begin{equation}
\Delta \phi = \phi_+ - \phi_m \approx 845 \quad {\rm and} \quad\Delta U \approx 7.88 \times 10^8\,,
\label{eq:Delta-phi-U}
\end{equation}
quantities which will be used later.
In the vicinity of the false vacuum, $\phi$ is approximately a free massive scalar field of mass $m$, and $U(\phi)$ has the form
\begin{equation}
U(\phi) \approx U(\phi_+) + \frac{1}{2} m^2 \, (\phi - \phi_+)^2 +O[(\phi - \phi_+)^3]\,.
\end{equation}
For small $\epsilon$, we have the estimate
\begin{equation}
 m \approx a \, \sqrt{\lambda} \,.
 \label{eq:m-est}
 \end{equation}
For the parameters given in  Eq.~\eqref{eq:parameters}, a more precise value is
 \begin{equation}
 m \approx 91.7267 \,.
 \label{eq:m-value}
 \end{equation}

If the system is initially in the false vacuum state, we expect it to be unstable to decay to the true vacuum. In principle, the decay probability may be computed
in the path integral formalism as
\begin{align}
P\sim\sum\limits_{\phi(x)}\exp(-S_E[\phi]),
\label{summation}
\end{align}
where $\phi(x)$ is a field configuration in Euclidean space which approaches both $\phi_+$ and
$\phi_-$ in different limits. Here $S_E$ is the Euclidean action, given by
\begin{align}
 S_E[\phi]=\int dt_E d\vec x \left[\frac12 \left(\frac{\partial\phi}{\partial
 t_E}\right)^2+\frac12(\vec\nabla\phi)^2+U\right]\,,
 \label{action}
 \end{align}
where the Wick rotation $t_E = i t$ has been performed.  The summation in Eq.~\eqref{summation} requires a sum over configurations $\phi(x)$,
which cannot be computed exactly by any known methods. The instanton approximation assumes that this sum is dominated by
solutions of the Euclidean version of Eq.~\eqref{eq:eom} near that with the
lowest Euclidean action. Coleman calls this the ``bounce" solution, for which $S_E =B$. In the instanton approximation, the decay probability becomes
\begin{equation}
P \propto  \,{\rm e}^{-B}\,,
\label{eq:bounce probability}
\end{equation}
or specifically we can write the decay rate per unit volume as
\begin{equation}
\Gamma_I \approx K \,{\rm e}^{-B}\,.
\label{eq:bounce rate}
\end{equation}
The prefactor $K$ is treated by Callan and Coleman~\cite{CC77}, who show that it may be expressed as a functional determinant which has the dimensions
of the reciprocal of the product of time and volume, so that Eq.~(\ref{eq:bounce rate}) may be interpreted as a rate per unit spatial volume.



\begin{figure}[htbp]
\includegraphics[scale=0.6]{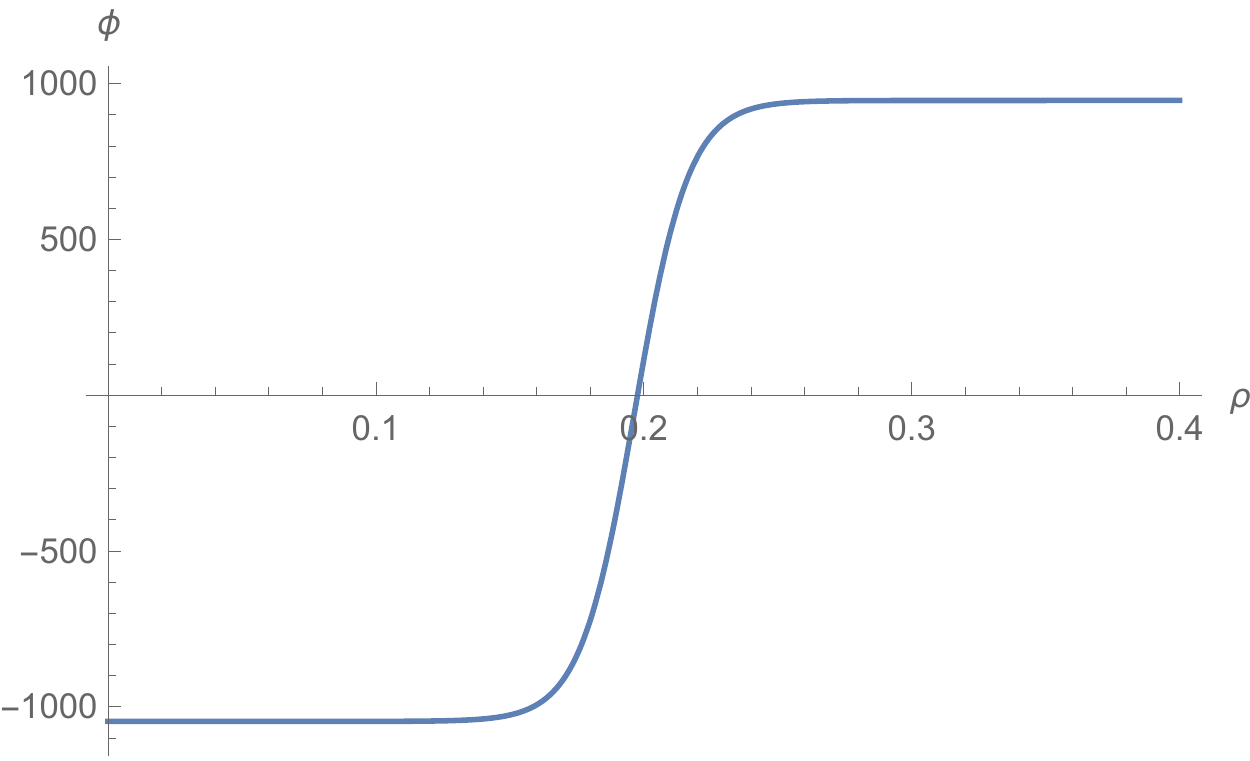}
\caption{The bounce solution $\phi(\rho)$ for the potential given in Eq.~\eqref{eq:potential} with the choice of parameters in Eq.~\eqref{eq:parameters}.
Here $\phi(0) = \phi_-$ and $\phi(\rho) \rightarrow \phi_+$ as $\rho \rightarrow \infty$. }
\label{fig:bouncesolution}
\end{figure}

The bounce solution is assumed to be $O(4)$ symmetric, and hence
Eq.~\eqref{eq:eom} becomes an ordinary differential equation with independent variable $\rho = (t_E^2 + |{\bf x|}^2 )^{1/2}$, the radius in four-dimensional
Euclidean space. Here $t_E = i t$ is the Euclidean time. This equation can be integrated numerically. We do this using the software package described in Ref.~\cite{MOS17}.
The result for the case of the parameters given in Eq.~\eqref{eq:parameters} is illustrated in
Fig.~\ref{fig:bouncesolution}.
This solution describes the nucleation of a bubble filled with the true vacuum, which nucleates in the false vacuum with an initial radius of $\rho = \rho_0 \approx 0.2$.
The bounce action in this case is
\begin{equation}
B \approx 2.47 \times 10^6 \,.
\label{eq:B_num}
\end{equation}
The function $\phi(\rho)$ describes the spatial configuration of the bubble when it nucleates, with $\phi$ varying from the false vacuum value in the interior of the
bubble to the true vacuum value on the exterior. The wall of the bubble is the region where $\phi$ varies most rapidly with increasing $\rho$. After nucleation,
the bubble expands, with $\rho$ taking the Lorentzian form, $\rho =  ( |{\bf x|}^2 -t^2 )^{1/2}$. Thus the expansion of the wall of the bubble is described by a
spacetime hyperbola, which is the world line of a uniformly accelerated particle. The expansion requires that the volume energy on the interior of the bubble at
least balance the surface energy in the wall. As the bubble expands, the volume energy decreases due to the lower energy density of the true vacuum relative to
the false vacuum, while the surface energy arising from spatial gradients in the wall increases. Bubbles with radii less than the minimum value, $\rho_0$, will
collapse rather than expand. This is similar to nucleation of bubbles of vapor in a boiling liquid.

There is one limit in which the bounce solution may be obtained analytically, when $\epsilon \ll 1$. This case, called the thin-wall approximation, is treated in
Sec.~IV of Ref.~\cite{C77}. However, there seem to be a few numerical errors in this reference. We find that there should be an additional factor of $2$ on the
right hand side of Eq.~(4.12), which leads to an additional factor of $16$ on the  right hand side of Eq.~(4.21) of Ref.~\cite{C77}. There also appear to be a factor of $2$
missing from the last term of Eq.~(4.15), but this is corrected in the following expression, Eq.~(4.16).
 Note that the parameter $\epsilon$ is defined differently in Ref.~\cite{C77} from our definition. There the potential is written in the form
 \begin{equation}
U(\phi)=\frac{\lambda}{8}(\phi^2-a^2)^2+\frac{\epsilon_1 }{2a}(\phi-a)\,,
\end{equation}
where $\epsilon_1 = \epsilon \, \lambda \, a^4$, when $\epsilon$ is defined as in Eq.~\eqref{eq:potential}. Accounting for both these corrections, and the change in
notation,  the initial radius of the bubble becomes
\begin{equation}
\rho_{0tw} = \frac{2}{ \epsilon \, a \,\sqrt{\lambda}} \,,
\label{eq:tw-radius}
\end{equation}
and the bounce action is
\begin{equation}
B_{tw} = \frac{8\,\pi^2}{3 \, \epsilon^3 \, \lambda} \,.
\label{eq:Btw}
\end{equation}

The above corrections seem to be consistent with the results of Garbrecht and Millington~\cite{GM15}, who write the potential in the form
\begin{equation}
U = \frac{\lambda_1}{4!}\, \Phi^4 + \frac{g}{6}\, \Phi^3 - \frac{1}{2} m_\Phi^2 \, \Phi^2 +U_0 \,.
\label{eq:U-GM}
\end{equation}
This form is equivalent to Eq.~(\ref{eq:potential}), as may be seen by letting $\Phi = \phi + s$ and expanding Eq.~(\ref{eq:U-GM}) to find
\begin{equation}
\lambda = \frac{1}{3}\, \lambda_1 \,,
\label{eq:GM-1}
\end{equation}

\begin{equation}
s = -\frac{g}{\lambda_1} \, ,
\label{eq:GM-2}
\end{equation}

\begin{equation}
a^2 = \frac{3}{\lambda_1^2} \, (2 m_\Phi^2 \lambda_1 +g^2) \approx \frac{6\, m_\Phi^2 }{\lambda_1}  +O(g^2) \,,
\label{eq:GM-3}
\end{equation}

and
\begin{equation}
\epsilon = 2 g\;  \frac{3 m_\Phi^2  \lambda_1 +g^2}{3^{3/2} (2 m_\Phi^2 \lambda_1 +g^2)^{3/2}}  \approx \frac{g}{\sqrt{6 \lambda_1} \, m_\Phi} +O(g^2) \,.
\label{eq:GM-4}
\end{equation}
Note in Eqs.~(\ref{eq:GM-3}) and (\ref{eq:GM-4}), we have given an expansion to first order in $g$, which reveals that $g \ll \sqrt{\lambda_1} \, m_\Phi$ is equivalent
to $\epsilon \ll 1$, the thin wall approximation. We may use the above expressions to show that  Eq.~\eqref{eq:tw-radius} and Eq.~\eqref{eq:Btw} are equivalent to
the corresponding results given by Garbrecht and Millington in Eqs.~(12) and (13), respectively, of  Ref.~\cite{GM15}.

As noted above, Callan and Coleman~\cite{CC77} express the prefactor $K$ in Eq.~(\ref{eq:bounce rate}) as a functional integral which is difficult to
calculate explicitly in general. However, it may be found explicitly in the thin wall approximation from a functional integration over zero modes, which was done
by subsequent authors, including Garbrecht and Millington~\cite{GM15}. The result in the latter reference may be expressed as
\begin{equation}
K \approx \frac{32 \pi^2}{9\sqrt{3}\;} \epsilon^{-7} \, a^4 \,     [1 + O(\epsilon^2)]\,.
\label{eq:K-tw}
\end{equation}
Recall that $\epsilon$ and $\lambda$ are dimensionless constants, but that $a$ has dimensions of $1/{\rm time} = 1/{\rm length}$, so $K$ has the correct
dimensions for a rate per unit volume.
Note that $K\propto \epsilon^{-7}$ is very large in the thin wall limit, but the decay rate given by Eq.~(\ref{eq:bounce rate}) vanishes as $\epsilon \rightarrow 0$
due to the fact that $B_{tw} \propto \epsilon^{-3}$ appears in the exponential.




 \section{Classical Field Dynamics}
 \label{sec:class}

 \subsection{Energy Conservation}
 \label{sec:energy}

 Consider a real classical field $\phi$ which satisfies the equation of motion Eq.~\eqref{eq:eom} in Minkowski spacetime, so
 \begin{equation}
\Box \phi = - \partial_t^2\phi + \nabla^2 \phi = U'(\phi) \,.
\label{eq:eom2}
\end{equation}
The associated energy density of this field is
\begin{equation}
T_{tt} = \frac{1}{2} [ (\partial_t \phi)^2 + |\mathbf{\nabla} \phi|^2 ] + U(\phi) \,.
\label{eq:density}
\end{equation}
 Consider a finite spatial region $R$, and define the field energy in this region at time $t$ by
 \begin{equation}
E_R(t) = \int_R d^3x \, T_{tt}({\bf x},t)\,.
\label{eq:E-R}
\end{equation}
Then
\begin{equation}
\frac{d E_R}{dt} =  \int_R d^3x \,  \{ \partial_t \phi\, [\partial_t^2 \phi + U'(\phi)] + \mathbf{\nabla} \phi \cdot  \mathbf{\nabla} \partial_t \phi \}
=  \int_R d^3x \,  \{ \partial_t \phi\,  [\partial_t^2 \phi - \nabla^2 \phi  + U'(\phi)] + \mathbf{\nabla} \cdot  (\partial_t \phi \, \mathbf{\nabla} \phi) \} \,.
\label{eq:Edot}
\end{equation}
The last term in the above expression is a total divergence, which may be written as a surface integral over the boundary $S$ of region $R$:
\begin{equation}
\int_R d^3x \,  \mathbf{\nabla} \cdot  (\partial_t \phi \, \mathbf{\nabla} \phi) = \oint_S d{\bf S} \cdot (\partial_t \phi \, \mathbf{\nabla} \phi) \,,
\end{equation}
which will vanish if either $\partial_t \phi$ or the normal component of $ \mathbf{\nabla} \phi$ vanish on each point of $S$.
In this case, Eqs.~\eqref{eq:eom2} and \eqref{eq:Edot} imply that $E_R$ is a constant.

In the case of a spatially homogeneous field, $\phi = \phi(t)$, the spatial integration simply produces a constant factor of the volume
of $R$, and $E_R \propto (\dot \phi)^2/2 +U(\phi(t))$. This is of the same form as the energy of a point particle in a potential.
The turning points of the motion occur when $U(\phi)$ reaches its maximum value, and $\dot \phi =0$.
The case of an inhomogeneous field is somewhat
more complicated, but one can still approximately identify turning points as occurring when $\int_R d^3x \,  U(\phi(t)) \approx E_R$,
when the contributions of both   $(\dot \phi)^2$ and $ |\mathbf{\nabla} \phi|^2$ to $E_R$ are small.  Thus if the variation of $\phi$
within $R$ is small compared to the variation between a pair of turning points of $ U(\phi)$, we can view $R$ as a localized region
which moves between these turning points much as does a point particle.

\subsection{Motion over a Barrier}
\label{sec:fly-over}

Consider the dynamics of a finite region in a potential with local minima, such as illustrated in  Fig.~\ref{fig:potential}. If the energy  $E_R$
of the region is small, then the motion is expected to be confined to be near one minimum, analogous to that of a classical particle
oscillating about a minimum. However, larger values of  $E_R$ might allow the region to pass over the  local maximum. One way to
achieve this would be to impose initial conditions that  $\phi = \phi_+$, so the system starts in the false vacuum, and
$\dot \phi= \dot \phi_0 \not=0$ in a finite spatial region $R$.
If  $|\dot \phi_0|$ and the size of  $R$ are sufficiently large, then this region can move over the local maximum at   $\phi_m$ to the
global minimum at  $\phi = \phi_-$.   In the potential  illustrated in  Fig.~\ref{fig:potential}, this can happen more easily if $\dot \phi_0 < 0$.
This would be a classical version of quantum false vacuum decay. The region $R$ becomes a bubble
similar to those discussed  in Sect.~\ref{sec:instanton}. Again, there is a competition between the volume energy inside the bubble
and surface energy in the wall, which requires that the bubble have a minimum size before it can expand rather than collapse. If
the bubble does expand, it eventually fills all of space with a region where $\phi \approx \phi_-$, the global minimum. This will be
illustrated in some numerical simulations in Sect.~\ref{sec:num}.  In the next section, we discuss a model where the classical initial condition on
$\dot \phi_0$ is replaced by the effect of a quantum field fluctuation.

\section{Vacuum Decay Induced by ${\dot\phi}$ Fluctuations}
\label{sec:phidot}

In this section, we will be concerned with possible effects of the quantum fluctuations of the scalar field,  $\phi$, and its space and time
derivatives, such as  $\dot\phi$.  We assume that initially $\phi \approx \phi_+$, the false vacuum value, and that to leading order, $\phi$ is a
linear quantum field with mass $m$.

\subsection{The Probability Distribution for Spacetime Averaged $\dot\phi$ Fluctuations}
\label{sec:phidot-distribution}

The fluctuations of a field operator at a single spacetime point are not meaningful, but averages over space and/or time are well defined.
We can view these averages as the result of a measurement of the field in a finite region. Here we consider averaging over both  space and
time regions and write the average of $\dot\phi$ as
\begin{align}
  \overline{\dot\phi}=\int dt~f(t)\int d^3 x~g({\bf x})~\dot\phi({\bf x}, t) \,,
  \label{overlinephidot}
\end{align}
where the averaging functions are normalized by
\begin{equation}
\int dt~f(t) = \int d^3 x~g({\bf x}) =1\,.
\label{eq:normalization}
\end{equation}
The probability distribution for $ \overline{\dot\phi}$ fluctuations in the vacuum state is a Gaussian function,
  \begin{align}
    P(\overline{\dot{\phi}})=\frac{1}{\sqrt{2\pi\sigma^2}}\exp\left(-\frac{{\overline{\dot \phi}}^2}{2\,
    \sigma^2}\right),
    \label{eq:pphidot}
  \end{align}
  where the variance $\sigma^2$ is
  \begin{align}
    \sigma^2=\langle  \overline{\dot\phi}^2\rangle
    =&\langle 0| \int dt_1 d^3 x_1~f(t_1)g({\bf x}_1)~{\dot\phi}(t_1,{\bf x}_1)\int dt_2~
    d^3 x_2~f(t_2)g({\bf x}_2)~{\dot\phi}(t_2,{\bf x}_2)|0\rangle \,.
\label{eq:sigma}
    \end{align}

We may write the linear operator $\dot\phi$ as
\begin{align}
\dot\phi(x)=\sum\limits_{\bf k} \sqrt{\frac{\omega}{2V}}\, [e^{i({\bf k}\cdot {\bf x} -\omega t)}a_k+e^{-i({\bf k}\cdot{\bf x} -\omega t)}a_k^\dagger]\,,
\label{dotphiexpand}
\end{align}
where  $V$ is a quantization volume, and  $\omega^2= k^2+m^2$.  In the large volume limit, this leads to an expression for the variance:
\begin{align}
\sigma^2
=& \frac{1}{2(2\pi)^3}\int d^3 k~\omega~\hat{g}^2({\bf k)} \hat{f}^2(\omega) \,,
\end{align}
where $\hat f(\omega )$ and $\hat g({\bf k})$ are  the Fourier transforms of the time and space sampling functions, respectively.
We assume that $\tau \not=0$ is the characteristic width of  $f(t)$, and hence is the time sampling interval. Similarly, let $\ell \not=0$ be the
characteristic spatial sampling interval. It is convenient to define dimensionless functions with unit sampling intervals by
\begin{align}
f_1(t)=\frac{1}{\tau}f\left(\frac {t}{\tau}\right),~
g_1({\bf x})=\frac{1}{\ell^3}g\left(\frac{ {\bf x}}{\ell}\right) \,.
 \label{unitsampling}
\end{align}
Let $\hat f_1$ and $\hat g_1$ be their corresponding Fourier transforms.
We can now express the variance as
\begin{align}
\sigma^2=\frac{\eta}{2\ell^3\tau}\,,
\label{eq:sigma2}
\end{align}
where
\begin{align}
\eta:=\frac{1}{(2\pi)^3}\int d^3 \kappa~\Omega~\hat{g}_1^2({\boldsymbol \kappa})
\hat{f}_1^2(\Omega) \,,
\label{eq:eta}
\end{align}
with ${\Omega}^2=\kappa^2\frac{\tau^2}{\ell^2}+m^2\tau^2$.
Note that $\eta$ is a dimensionless quantity which depends upon the functional forms of the sampling functions, as well as any two of the three
dimensionless variables $m \tau$, $m\ell$ and $\ell/\tau$.

In the limit where the mass vanishes, $m = 0$, we are left with two parameters, $\ell$ and $\tau$. However, the variance $\sigma^2$, will
be finite with either time averaging alone or spatial averaging alone. Thus we expect to have
\begin{equation}
\sigma^2 \propto \tau^{-4} \, , \qquad  \tau \agt \ell \,,
\label{eq:var-t-ave}
\end{equation}
or
\begin{equation}
\sigma^2 \propto \ell^{-4} \, , \qquad  \ell \agt \tau \,.
\label{eq:var-space-ave}
\end{equation}
Both of these limits are consistent with Eq.~\eqref{eq:sigma2}. When $\tau \agt \ell$, we expect $\eta \propto (\ell/\tau)^3$, but that
$\eta \propto \tau/\ell$ when   $\ell \agt \tau$.

Often we are interested in the probability of a fluctuation which exceeds a given threshold. Let $P(x)$ be a probability distribution, so
that $\int_{x_0}^{x_1} P(x) \, dx$ is the probability of finding $x_0 \leq x \leq x_1$ in a given measurement. Define the complementary
cumulative probability  by
\begin{equation}
P_>(y) = \int_y^\infty P(x) \, dx\,.
\label{eq:P>}
\end{equation}
This is the probability of finding $x \geq y$ in a given measurement. In the case where $P(x)$ is a Gaussian, such as given in Eq.~\eqref{eq:pphidot}
with $x =  \overline{\dot{\phi}}$,  $P_>(y)$ is expressible as an error function.
For large argument, it has the asymptotic form
\begin{equation}
P_>(y)  \sim \frac{\sigma}{\sqrt{2 \pi}} \, \exp\left[-\frac{y^2}{2 \sigma^2}\right] \; \left(\frac{1}{y} + O(y^{-2})  \right) \approx
 \exp\left[-\frac{y^2}{2 \sigma^2} - \ln(\sqrt{2 \pi} y/\sigma) \right] \,.
 \label{eq:P>asy}
\end{equation}
Thus if $y$ is large enough that the logarithm term may be neglected, then $P_>(y)$ has approximately the same functional form as does
$P(x)$.

\subsection{Compactly supported functions}
\label{sec:compact}

We adopt the view that the sampling functions $f(t)$ and $g({\bf x})$
should have compact support, meaning that they are strictly equal to zero outside of finite intervals. The physical motivation for
 this is that these functions should describe measurements made within finite time intervals and spatial regions. A temporal
 sampling function such as a Gaussian or Lorentzian has tails which extend into the past and the future, and strictly describes a measurement
 which began in the infinite past and continues into the infinite future. A better choice is an infinitely differentiable function with
 compact support. Such functions have Fourier transforms which fall faster than any power, but more slowly than an exponential
 function. A class of these functions was discussed in Refs~\cite{FF15} and \cite{FF19}, and have Fourier transforms with the
asymptotic forms
\begin{equation}
\hat f(\omega)\sim \exp[-(\omega \tau)^{\alpha}]  \,, \quad \omega \tau \gg 1 \,, \qquad \hat g(k) \sim \exp[-(k \ell)^{\lambda}] \,,  \quad k \ell \gg 1 \,,
\label{eq:asyFourier}
\end{equation}
where $\alpha$ and $\lambda$ are real constants which satisfy $0< \lambda \leq \alpha <1$. Here the spatial sampling function is
assumed to be spherically symmetric, so $g=g(r)$ and $\hat g = \hat g(k)$. The coordinate space switch-on or switch-off behavior is
linked to the values of $\alpha$ and $\lambda$. For example, if $f(t)$ switches on at $t = 0$, then it might have the form
\begin{equation}
f(t) \sim D\, t^{-\mu}\, \exp(-w\, t^{-\nu})
\end{equation}
as $t \rightarrow 0^+$ for some constants $D$, $\mu$, $w$, and $\nu$. The most important of these is $\nu$, which is related to the
parameter $\alpha$ in $\hat f$ by $\nu = \alpha/(1-\alpha)$. The choice $\alpha = 1/2$, where $\nu =1$, has special physical interest, as there
is an electrical circuit which switches on with this behavior~\cite{FF15}. Some explicit examples of compactly supported functions were
given in Sect.~IIB of Ref.~\cite{FF15} and in Appendix A of Ref.~\cite{FF19}.

Another choice can be given in coordinate space by
\begin{equation}
f_1(t)=C_f\left\{
\begin{array}{ll}
\exp\left(-\frac{1}{1-t^2}\right), &t\in[-1,1]\\
0, &t<-1~$or$~ t>1
\end{array}\label{samplingfunctiontime}
\right.
\end{equation}
and
\begin{equation}
g_1(\vec x)=C_g\left\{
\begin{array}{ll}
\exp\left(-\frac{1}{1-|\vec x|^2}\right), &|\vec x|\in[0,1]\\
0, & |\vec x|>1 \,,
\end{array}
\label{samplingfunctionspace}
\right.
\end{equation}
where $C_f$ and $C_g$ are normalization factors chosen so that  Eq.~\eqref{eq:normalization} holds.  The functions above
correspond to $\alpha =  \lambda =1/2$, and $\tau = \ell =1$.

\subsection{Effects of Large  $\overline{\dot\phi}$ Fluctuations}
\label{sec:effects}

Let us return to the process described in Sect.~\ref{sec:fly-over}, where a classical initial condition on  ${\dot\phi}$ could cause
a finite spatial region to move from the false vacuum, over the barrier to the true vacuum. However, we now consider quantum
vacuum fluctuation of a space and time average, $\overline{\dot\phi}$, and ask under what conditions it might produce the same
effect. The probability of a sufficiently large fluctuation may be estimated from Eqs.~\eqref{eq:pphidot} and \eqref{eq:sigma2} once we
have estimates for $\overline{\dot\phi}$, $\tau$, $\ell$ and $\eta$. We expect that we need $\frac{1}{2} {\overline{\dot\phi}}^2 \agt  \Delta U$,
where
\begin{equation}
 \Delta U = U(\phi_0)-U(\phi_F)
 \label{eq:delta-U}
\end{equation}
is the height of the potential barrier above the false vacuum level. Take the minimum value of the magnitude of $\overline{\dot\phi}$ to be
\begin{equation}
|\dot\phi_0| = (2\Delta U)^{\frac12}\,.
\label{eq:phi-dot}
\end{equation}
Let the minimum value of  $\tau$ be
\begin{equation}
\tau_0 = \frac{\Delta \phi}{|\dot\phi_0|}\,,
\label{eq:tau-0}
\end{equation}
where $\Delta \phi = |\phi_0-\phi_F|$. That is, $\tau_0$ is the time that would be required  for the field to change by $\Delta \phi$ if it maintained
an average speed of  $\dot\phi_0$. Finally, we estimate that the minimum size of the spatial averaging region should be of the order of the radius at which
a bubble could nucleate in the given potential, in the thin wall approximation,
\begin{equation}
\ell_0= \rho_{0tw}\,.
\label{eq:ell0}
\end{equation}
Recall that this is the minimum radius at which the internal pressure can balance the tension in the bubble wall.
More generally, we expect the values of $\dot\phi$, $\tau$, and $\ell$ to be of the order of the minimum values estimated above. Set
\begin{equation}
\dot\phi = V\, \dot\phi_0 \, , \quad     \tau = T\, \tau_0\, , \quad  \ell = L\, \ell_0\,,
\label{eq:scaling}
\end{equation}
where $V$, $T$, and $L$ are constants of order unity.

Let
\begin{equation}
A=\frac{{\overline{\dot\phi}}^2 \ell^3\tau}{\eta}  =   V^2 \, L^3\, T \; \frac{{\overline{\dot\phi_0}}^2 \ell_0^3\tau_0}{\eta}  \,.
\label{eq:A-def}
\end{equation}
The probability of a  $\overline{\dot\phi}$-fluctuation which is  sufficiently large to move a region over the potential barrier is of order
\begin{equation}
P(\overline{\dot{\phi}}) \approx P_>(\overline{\dot{\phi}}) \approx  {\rm e}^{-A}\,.
\label{eq:PA}
\end{equation}
This fluctuation occurs in spatial volume of order $\ell^3$ on a timescale of about $\tau$, so the corresponding rate per unit volume of vacuum
decay by this mechanism is of order
\begin{equation}
\Gamma_{\overline{\dot{\phi}}}      \approx  \frac{1}{\tau \, \ell^3} \;  {\rm e}^{-A}\,.
\label{eq:rate1}
\end{equation}
This is to be compared with the result in the instanton approximation, Eq.~\eqref{eq:bounce rate}.


\subsubsection{Thin Wall Case}
\label{sec:thin}

Here we wish to compare estimates of $A$ with $B$ in the thin wall approximation. Take the example of the potential given in Eq.~\eqref{eq:potential}
for the case $\epsilon \ll 1$. To lowest order in $\epsilon$, we have
\begin{equation}
\phi_F \approx a, \qquad \phi_0 \approx 0,  \quad  {\rm and} \quad \Delta U \approx \frac{1}{8} \lambda \, a^4 \,,
\label{eq:tw-parameters}
\end{equation}
leading to
\begin{equation}
\dot\phi_0 \approx \frac{1}{2}\, \sqrt{\lambda}\, a^2\,,
\end{equation}
and
\begin{equation}
 \tau_0 \approx \frac{2}{\sqrt{\lambda}\, a}\,.
 \label{eq:tau0}
\end{equation}
Note that $\tau_0 \approx 2/m$, where $m$ given in Eq.~\eqref{eq:m-est}, is the mass associated with the false vacuum state, and
hence   $\tau_0$  is of the order of the period of harmonic field oscillations  in the false vacuum.
If we combine these estimates with $\ell_0 \approx \rho_{0tw}$ and Eq.~\eqref{eq:tw-radius}, we find
\begin{equation}
A \approx \frac{4\, V^2 \, L^3\, T }{ \, \eta\, \epsilon^3 \, \lambda}\,.
\label{eq:Atw}
\end{equation}
Comparison with Eq.~\eqref{eq:Btw} shows that $A < B$, and hence ${\rm e}^{-A} > {\rm e}^{-B}$, if
\begin{equation}
\frac{\eta}{V^2 \, L^3\, T } > \frac{3}{2\,\pi^2} \approx 0.15 \,.
\label{eq:eta-bound}
\end{equation}
The value of $\eta$ depends upon the choices of the sampling functions $f(t)$ and $g({\bf x})$, but can be expected to be of order one, in which case
Eq.~\eqref{eq:eta-bound} will be satisfied if $V^2 \, L^3\, T$ is not too large.

However, comparison of the decay rate due to instanton effects, $\Gamma_I$ and that due to $\overline{\dot\phi}$ fluctuations, $\Gamma_{\overline{\dot{\phi}}} $,
also requires comparisons of the prefactors to the exponentials. Consider the case $T = L =1$ and use Eqs.~\eqref{eq:tw-radius}, \eqref{eq:ell0},  and
\eqref{eq:tau0} to find the $\Gamma_{\overline{\dot{\phi}}}$ prefactor to be
\begin{equation}
   \frac{1}{\tau \, \ell^3}  \approx    \frac{1}{16}\,   \epsilon^3 \,  \lambda^2 \, a^4 \,   \,.
\label{eq:ell3-tau}
\end{equation}
By contrast, the prefactor $K$, given in Eq.~\eqref{eq:K-tw} is larger by a factor of order $\epsilon^{-10}\lambda^{-2}$, so instanton effects will dominate if $A=B$. However,
if  $A < B$, a relatively small fractional difference can cause $\overline{\dot\phi}$ fluctuations to dominate. This will be illustrated in the next subsection.

\subsubsection{Numerical Simulations}
\label{sec:num}

Here we describe some numerical integrations of Eq.~\eqref{eq:eom2} with $U(\phi)$ as given in Eqs.~\eqref{eq:potential} and \eqref{eq:parameters}.
The initial condition is that $\phi(0) = \phi_+$, the false vacuum value, everywhere and that $\dot\phi(0)$ is a negative constant within a sphere of initial radius $\ell$. The
numerical solution, $\phi(t,r)$ is inspected to check that at least the region near $r=0$ has passed through $\phi = \phi_m$ and reached
$\phi \approx \phi_-$. If so, this describes a bubble of true vacuum surrounded by false vacuum formed by a $\overline{\dot\phi}$ fluctuation. However, a
quantum fluctuation is transient and its energy must be given up on some timescale $\tau = T \tau_0$. We model this effect by stopping the numerical
integration at $t=\tau$, and then restarting it with a new initial condition that $\phi(t=\tau,r)$ has the value found in the previous part of the integration,
but   $\dot\phi(t=\tau,r) =0$. That is, the value of $\dot\phi$ is set to zero for the beginning of the second part of the simulation. Equation~\eqref{eq:eom2}
is now further integrated with these initial conditions at $t=\tau$ to see if the bubble continues to expand.

\begin{figure}[htbp]
\begin{minipage}[t]{0.45\textwidth}
\includegraphics[width=1.2\textwidth,center]{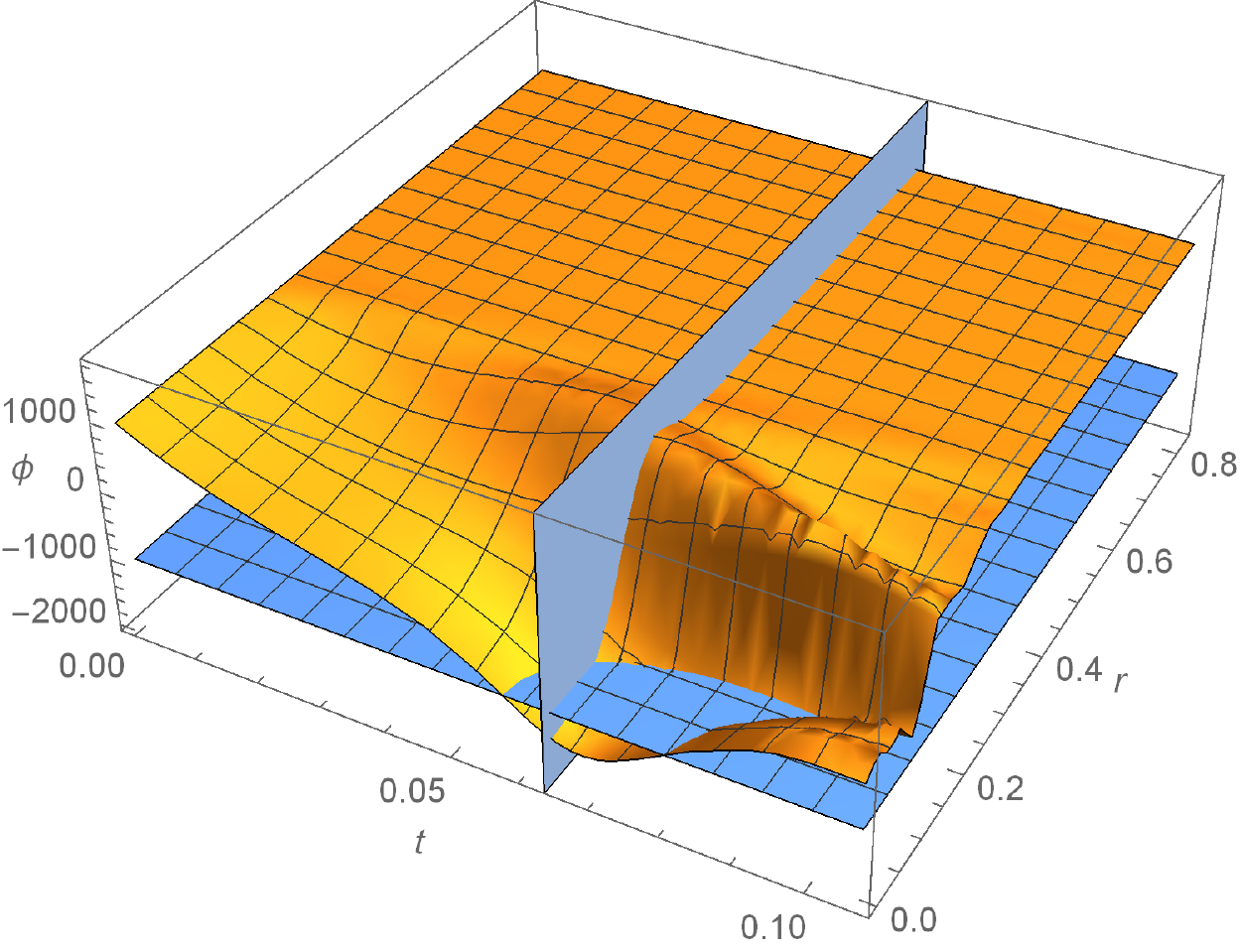}
\caption{The first part of the simulation,\\
for the fluctuation in the case $L=1.5$,\\ $T=1.25,~ V=1.5$. }
\label{simulation1}
\end{minipage}
\begin{minipage}[t]{0.45\textwidth}
\includegraphics[width=10cm,height=15cm]{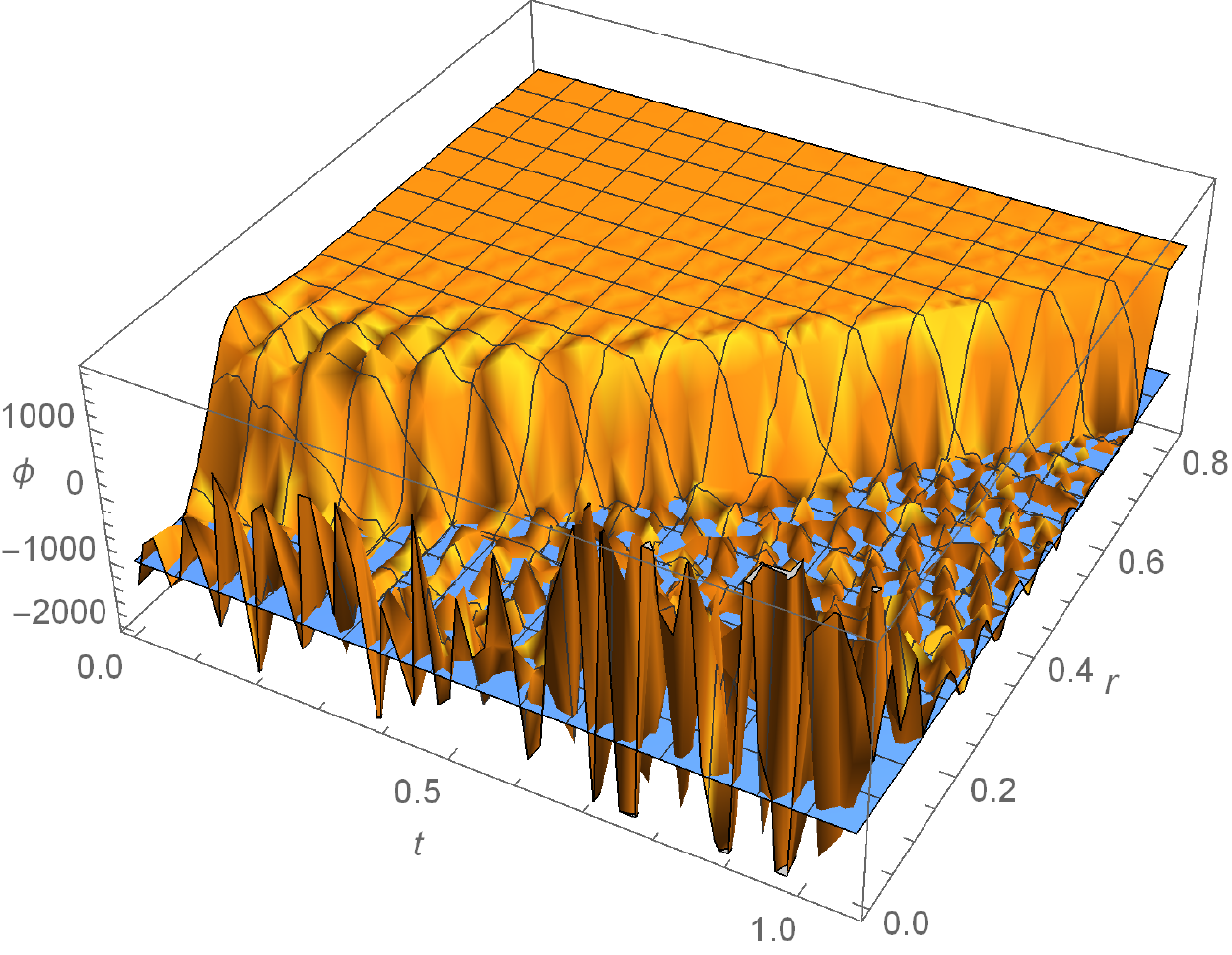}
\caption{The succeeding second part of the simulation, following Fig.~\ref{simulation1}.}
\label{simulation2}
\end{minipage}
\end{figure}

The first part of the simulation for the case  $L=1.5$, $T=1.25$, and $V=1.5$   is illustrated in Fig.~\ref{simulation1}, and the second part in Fig.~\ref{simulation2}.
The vertical plane in  Fig.~\ref{simulation1} corresponds to $t = \tau = 1.25\tau_0$, when the first part ends. By this time, the center of the bubble is in the true
vacuum phase. Figure~\ref{simulation2} illustrates the bubble expanding at close to the speed of light, with true vacuum in the interior, and false vacuum on the
exterior.

\begin{figure}[htbp]
\includegraphics[scale=0.5]{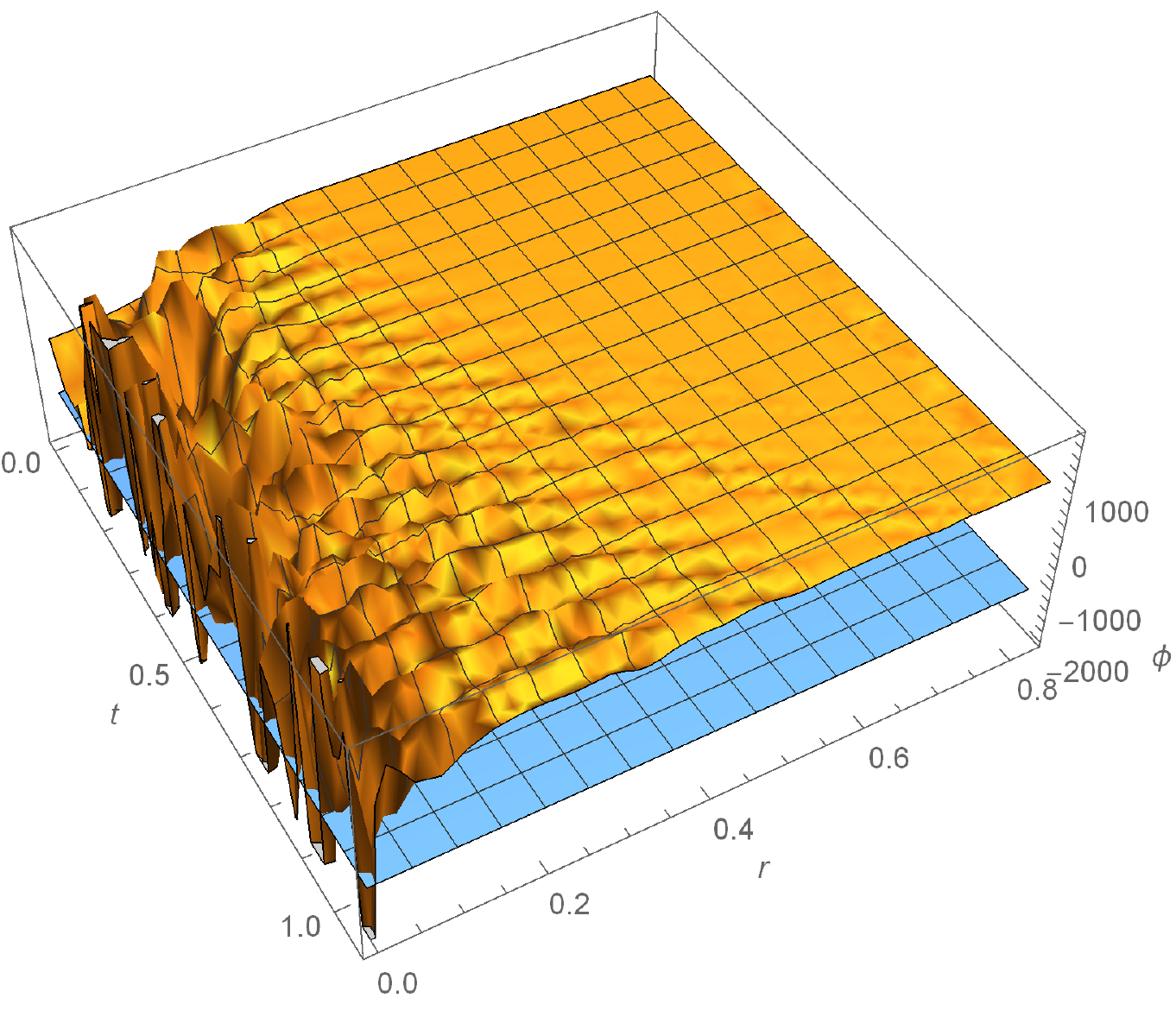}
\caption{The second step of the simulation for the case $L = 0.8; T= 1.25; V = 1.5$. The region of true vacuum fails to expand out.}
\label{fig:fail-to-expand}
\end{figure}

For some choices of the parameters, the bubble fails to expand in the second part of the simulation, but rather collapses. Here the internal pressure due to the true vacuum
energy is unable to overcome the wall tension. One such case is illustrated in Fig.~\ref{fig:fail-to-expand}.

Table~ \ref{tableA/B0.1} lists the results found with several choices of the parameters $L$, $T$, and $V$, along with the values of $A$ and $A/B$.  Here we have taken the
temporal and spatial sampling functions to be given by Eqs.~\eqref{samplingfunctiontime} and \eqref{samplingfunctionspace} , respectively, and then numerically computed
their Fourier transforms. These results are used in Eq.~\eqref{eq:eta} to find  value of $\eta$, which in turn is used to find $A$ from Eq.~\eqref{eq:A-def}. We also use
Eqs.~\eqref{eq:parameters}, \eqref{eq:Delta-phi-U}, \eqref{eq:phi-dot}, and \eqref{eq:tau-0}, and  $\ell_0 =0.2$. Note that the thin wall approximation is not used here in the
calculations of $B$.

In general, $A/B$ is roughly of order unity, indicating roughly comparable contributions from $\overline{\dot\phi}$ fluctuations and from quantum tunneling in the instanton
approximation to $\ln \Gamma$,
the logarithm of the decay rate. The ratio of the prefactors in $\Gamma_I$ and $\Gamma_{\overline{\dot{\phi}}}$ is of order
\begin{equation}
\tau \ell^3 \, K = \frac{2^{9}\,\pi^2}{9 \sqrt{3}}\, \epsilon^{-10}\lambda^{-2}\    \approx     {\rm e}^{5.78 - 10\, \ln \epsilon-2\,\ln \lambda}  \,,
\label{eq:prefactor-ratio}
\end{equation}
where we have used Eqs.~\eqref{eq:K-tw}  and \eqref{eq:ell3-tau}.
In the present calculations, where $\epsilon = 0.1$ and $\lambda=0.01$, this ratio becomes about 
$3.2 \times 10^{16} \approx {\rm e}^{38}$, so the ratio of the decay rates is
\begin{equation}
\frac{\Gamma_I}{\Gamma_{\overline{\dot{\phi}}}} \approx  {\rm e}^{A -B +38}\,.
\label{eq:rate-ratio1}
\end{equation}
 Although

However, although $A/B$ is of order one, the magnitudes of $A$ and $B$ are sufficiently large that $|A-B + 38| \approx |A -B| \gg 1$, so the ratio of rates in Eq.~\eqref{eq:rate-ratio1}
is either very large, as in the first four rows of Table~ \ref{tableA/B0.1}, or very small, as in the final six rows. In the former cases, quantum tunneling dominates, and the
latter,  $\overline{\dot\phi}$ fluctuations dominate.

 \begin{table}[htbp]
  	\caption{ The comparison between the two mechanisms of vacuum decay, for the potential barrier of
$\lambda = 0.01, a = 1000, \epsilon= 0.1$, under different fluctuations.}
  	  	\begin{center}
  		$\begin{array}{ccc|cccc|c|c}
 \text{L} & \text{T} & \text{V} & m\ell &  m\tau & \eta & \text{A} & \text{A/B}&\text{Expand out?} \\ \hline
 1.5 & 1.25 & 1.5 & 27.518 & 2.43893 & 0.431466 & 5.90252\times 10^6 & 2.38785 & \text{yes} \\
 1.4 & 1.25 & 1.5 & 25.6835 & 2.43893 & 0.430821 & 4.80616\times 10^6 & 1.94432 & \text{yes} \\
 1.3 & 1.25 & 1.5 & 23.8489 & 2.43893 & 0.43002 & 3.85524\times 10^6 & 1.55963 & \text{yes} \\
 1.2 & 1.25 & 1.5 & 22.0144 & 2.43893 & 0.429011 & 3.03938\times 10^6 & 1.22957 & \text{yes} \\
 1.1 & 1.25 & 1.5 & 20.1799 & 2.43893 & 0.427715 & 2.34819\times 10^6 & 0.949954 & \text{yes} \\
 1 & 1.25 & 1.5 & 18.3453 & 2.43893 & 0.426012 & 1.77128\times 10^6 & 0.716567 & \text{yes} \\
 0.9 & 1.25 & 1.5 & 16.5108 & 2.43893 & 0.423715 & 1.29827\times 10^6 & 0.52521 & \text{yes} \\
 0.8 & 1.25 & 1.5 & 14.6763 & 2.43893 & 0.420515 & 918753. & 0.371679 & \text{no} \\
 0.7 & 1.25 & 1.5 & 12.8417 & 2.43893 & 0.415878 & 622355. & 0.251772 & \text{no} \\
 0.6 & 1.25 & 1.5 & 11.0072 & 2.43893 & 0.408819 & 398688. & 0.161288 & \text{no} \\
\end{array}$
  	\end{center}
  \label{tableA/B0.1}
   \end{table}



 \subsection{Anti-Correlated Fluctuations}
\label{sec:anticorr}

As noted above, we expect that the energy borrowed by the classical field from the quantum vacuum will tend to be
returned on a timescale of order $\tau$. This corresponds to the end of the first part of the simulations described in
Sec.~\ref{sec:num} and illustrated in Fig.~\ref{simulation1}, and arises from the tendency of vacuum fluctuations to be
anti-correlated. This effect was discussed in Refs.~\cite{FR05,PF11}, where a correlation function was used to show
that a typical fluctuation of quantities such as energy density or electric field tends to be followed by a fluctuation with
the opposite sign. Here a typical fluctuation means one whose squared magnitude is of the order of the variance,
$\sigma^2$. This anti-correlation acts to enforce energy conservation in a free field theory on a long time scale.
Note from Eqs.~\eqref{eq:pphidot} and \eqref{eq:sigma2}, the probability $ P(\overline{\dot{\phi}})$ decreases rapidly
as $\tau$ increases for fixed $\overline{\dot{\phi}}$ and $\ell$. This arises from the anti-correlations, which make it more
difficult to observe a given value of $\overline{\dot{\phi}}$  over a longer averaging timescale. Note, however, that the
effect of the original fluctuation is not guaranteed to be exactly canceled on any finite time scale. The analysis
using correlation functions in Refs.~\cite{FR05,PF11} shows that on time scales a few times that associated with the
original fluctuation, cancellation is just somewhat more likely than non-cancellation.

This raises the question of whether an anti-fluctuation is likely to undo the effect of the initial large
$\overline{\dot{\phi}}$  fluctuation, and send the system back over the barrier to the false vacuum state after it has
reached the true vacuum state. We argue that such an  anti-fluctuation is unlikely. First, the arguments for anti-correlated
fluctuations given in Refs.~\cite{FR05,PF11} rely upon an operator correlation function, or two-point function. The large
fluctuations, large compared to the variance, are described by higher moments of the operator, or by $n$-point functions
with $n \gg1$. It is not clear if large fluctuations will soon be followed by equally large anti-fluctuations. Even if they are,
there would be a limited time window for the anti-fluctuation to return the system to the false vacuum state. Once a bubble
of true vacuum has formed and begun to expand rapidly, it is unlikely that an anti-fluctuation could stop this essentially
classical expansion. The most that one can expect of the anti-fluctuation is that it takes back the energy borrowed from the
quantum field by the original fluctuation. Once the bubble has begun to expand, its energy quickly becomes much larger than
this value, and the bubble becomes a classical field configuration.

\section{Vacuum Decay Induced by Quadratic Operator Fluctuations}
\label{sec:quad}

\subsection{Probability Distributions}
\label{sec:quad-distribution}

In the previous section, we discussed the possibility that vacuum fluctuations of a linear operator, such as the spacetime average of $\dot\phi$, could
induce decay of the false vacuum. Now we turn to the effects of the fluctuations of a quadratic operator, such as ${\dot\phi}^2$. The probability
distributions for such operators have been treated in Refs.~\cite{FFR12,FF15,SFF18,FF19,WFS21}.  Just as is the case
for $\dot\phi$, a quadratic operator must also be averaged in spacetime before a meaningful probability distribution can be defined. Recall that
for a linear operator, the averaging could be over only time or only space, although we selected averaging in both as being more physically realistic.
The averaging of a quadratic operator must be in time, as spatial averaging alone does not suffice. As before, we consider a spacetime average.
A key result is that the probability distribution for an averaged quadratic operator, such as $\overline{\dot\phi^2}$,  falls more slowly than exponentially
for large fluctuations. This means that large fluctuations of $\overline{\dot\phi^2}$ and similar operators are more likely than one might expect, and
hence may have larger physical effects than linear operator fluctuations.

Consider the case of a space and time average of normal ordered $:{\dot\phi}^2:$,
\begin{equation}
\overline{\dot\phi^2} = \int dt~f(t)\int d^3 x~g({\bf x})~:\dot\phi^2({\bf x}, t): \,,
\label{eq:STA}
\end{equation}
where we again take $f(t)$ and $g({\bf x})$ to be functions with compact support whose Fourier transforms have the asymptotic forms given in
Eq.~\eqref{eq:asyFourier}.
It is shown in Ref.~\cite{FF19} that  when $\lambda  \leq \alpha < 1$, the probability distribution and the complementary cumulative distribution functions
for $\overline{\dot\phi^2}$ have the asymptotic forms,
\begin{equation}
P(\overline{\dot\phi^2}) \sim   P_>(\overline{\dot\phi^2}) \sim   \exp[-a_1 (\tau^4 \, \overline{\dot\phi^2})^\alpha] \,,
\label{eq:P1}
\end{equation}
when $\ell \alt \tau$, and
\begin{equation}
P(\overline{\dot\phi^2}) \sim P_>(\overline{\dot\phi^2}) \sim  \exp[-a_2 (\ell^4 \, \overline{\dot\phi^2})^\alpha] \,,
\label{eq:P2}
\end{equation}
when $\ell \agt \tau$. Here $a_1$ and $a_2$ are constants of order unity. Note that both $\tau^4\, \overline{\dot\phi^2}$ and $\ell^4 \, \overline{\dot\phi^2}$
are  dimensionless measures of the magnitude of $\overline{\dot\phi^2}$. The above asymptotic forms hold when the arguments of the exponentials are large
compared to one. As in Sect.~\ref{sec:phidot-distribution}, we are assuming that logarithm terms inside the exponentials are subdominant.

 Note that it is the parameter $\alpha$ associated with $f(t)$ which governs the probability
of large fluctuations.  Because $\alpha < 1$, comparison of Eq.~\eqref{eq:pphidot} with either of Eq.~\eqref{eq:P1} or  Eq.~\eqref{eq:P2}
shows that the probability of a large $\overline{\dot\phi^2}$ fluctuation
can be much greater than that of the corresponding $\overline{\dot\phi}$ fluctuation for which ${\overline{\dot\phi}}^2 = \overline{\dot\phi^2}$.

The asymptotic probability distributions given in Eqs.~\eqref{eq:P1} and \eqref{eq:P2} apply to other quadratic operators with the same dimensions as
$\overline{\dot\phi^2}$, including components of the stress tensor. Large radiation pressure fluctuations were discussed in Ref.~\cite{HF16}, where it was argued that
they can sometimes give a significant contribution to the barrier penetration rate of quantum particles. However, this requires especially small values of the switching parameter, $\alpha\alt 1/3$. For larger values of $\alpha$, the radiation pressure fluctuation contribution is small compared to the barrier penetration rate found in the WKB approximation.
It is important to note that the results summarized here are for $3+1$ dimensions. A $1+1$ dimensional model was treated in Ref.~\cite{FFR10}, and the probability
distribution was found to fall more rapidly than in $3+1$ dimensions.

\subsection{Effects of Large  $\overline{\dot\phi^2}$  Fluctuations}
\label{sec:effects2}

We argued in Sect.~\ref{sec:effects} that the vacuum fluctuations of  $\overline{\dot\phi}$ can produce a contribution to false vacuum decay which is
comparable to the instanton contribution. Further, we have just seen that the probability of a $\overline{\dot\phi^2}$ fluctuation can be much larger
than that of a comparable $\overline{\dot\phi}$ fluctuation, for which $(\overline{\dot\phi})^2 \approx  \overline{\dot\phi^2}$. However, the possible effects
of the two types of fluctuations are potentially very different. In Sect.~\ref{sec:effects}, we treated the effect of a large $\overline{\dot\phi}$ fluctuation
as giving an initial condition for integration of the classical equation of motion for a self-coupled scalar field. We cannot expect that a large
$\overline{\dot\phi^2}$ fluctuation will generally be accompanied by a comparable $\overline{\dot\phi}$ fluctuation. It is more likely that a large
$\overline{\dot\phi^2}$ fluctuation will increase the energy of a finite region without necessarily changing the mean value of the classical scalar field
significantly. This is similar to thermal fluctuations, which satisfy a Boltzmann probability distribution and can be described by a sphaleron~\cite{KM84,AM87-88}.
Here a thermal fluctuation causes  a finite region to go over the potential barrier.

\begin{figure}[htbp]
\includegraphics[scale=0.2]{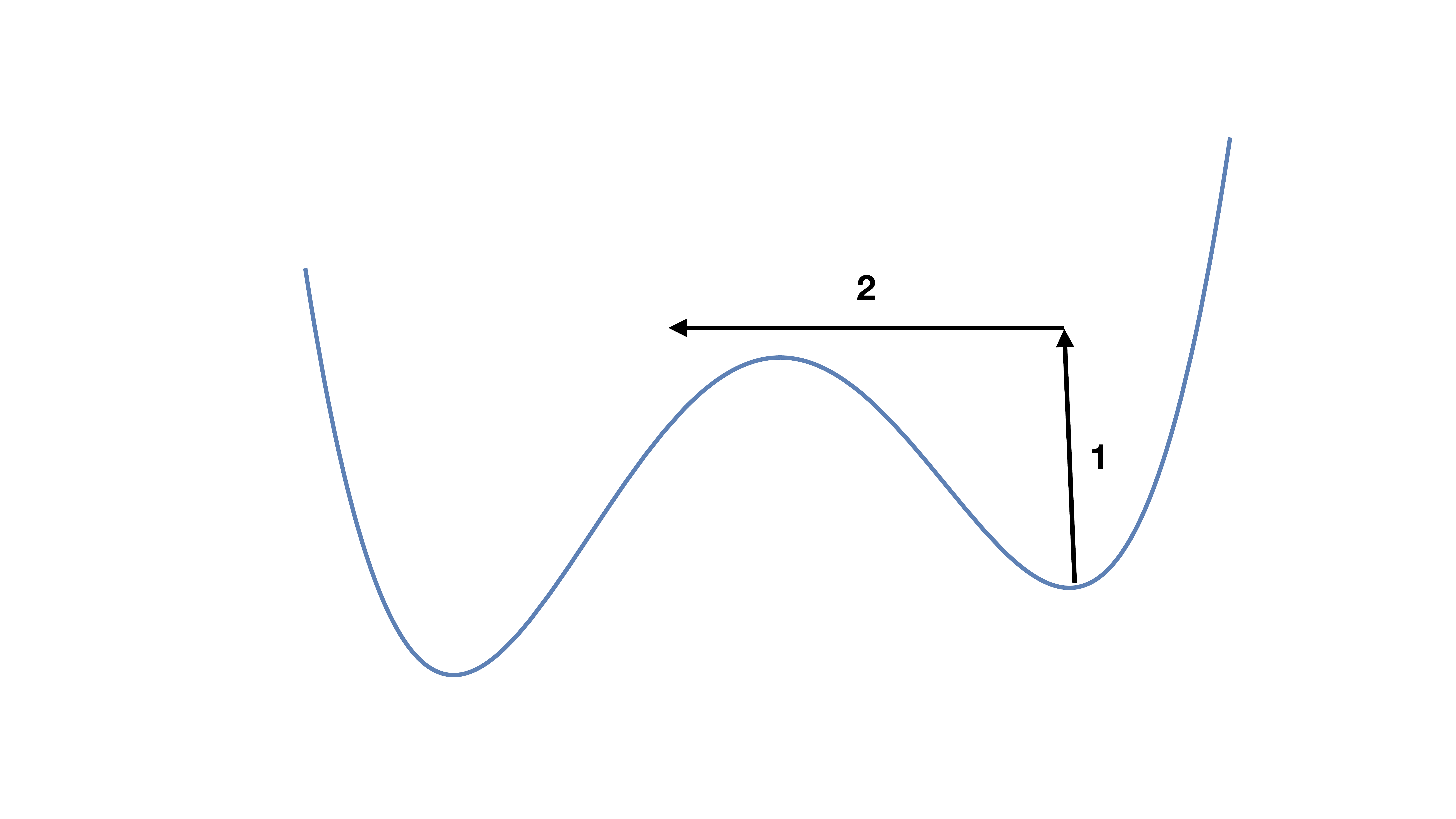}
\caption{Here the combined effects of a large $\overline{\dot\phi^2}$ fluctuation (Step 1), and a subsequent $\overline{\dot\phi}$ fluctuation (Step 2) are illustrated.}
\label{fig:2-step}
\end{figure}

However, if the mean value of the classical scalar field remains constant, then this region would seem to return to the false vacuum state when the
$\overline{\dot\phi^2}$ fluctuation has finished. One way to avoid this would be a simultaneous $\overline{\dot\phi}$ fluctuation which pushes the region over
the potential maximum, so that it ends in the true vacuum state.
The effects of both fluctuations are illustrated in Fig.~\ref{fig:2-step}. In principle, the time scales of the two fluctuations could be different, but here we assume that
they are of the same order.

The minimal $\overline{\dot\phi^2}$ fluctuation  needed raise the region above the potential maximum has a magnitude of $\overline{\dot\phi^2} \approx 2\, \Delta U$,
where $\Delta U$ is the potential difference defined in Eq.~\eqref{eq:delta-U}.  If $\ell \agt \tau$ and we set $a_2 \approx 1$, in Eq.~\eqref{eq:P2},  the probability of
the required $\overline{\dot\phi^2}$ fluctuation is of order
\begin{equation}
P_1(\overline{\dot\phi^2}) \approx P_>(\overline{\dot\phi^2}) \approx \exp[ -( 2\, \ell^4 \, \Delta U)^\alpha] \,.
\label{eq:Pstep1}
\end{equation}
Here the subscript $1$ refers to the first step in Fig.~\ref{fig:2-step}, and we assume that $( 2\, \ell^4 \, \Delta U)^\alpha \gg 1$, so that Eq.~\eqref{eq:P2} holds.
We may view a positive  $\overline{\dot\phi^2}$ fluctuation as raising the average energy density, and hence the energy $E_R$, defined in Eq.~\eqref{eq:E-R},
of the region.

The second step, denoted by the label $2$ in Fig.~\ref{fig:2-step}, involves a linear field fluctuation, similar to those treated in Sec.~\ref{sec:effects}. However,
now we may treat the field as approximately massless, and assume that Eq.~\eqref{eq:var-space-ave} holds if $\ell \agt \tau$. In this case, the probability
of a large $\overline{\dot\phi}$ fluctuation is approximately
\begin{equation}
P_2(\overline{\dot\phi}) \approx P_>(\overline{\dot\phi}) \approx \exp( -  \ell^4 \, {\overline{\dot\phi}}^2) \,.
\label{eq:Pstep2}
\end{equation}
However, unlike in Sec.~\ref{sec:effects}, now the $\overline{\dot\phi}$ fluctuation does not need to raise  a region from the false vacuum minimum over the
potential barrier, but rather simply has to translate by $\Delta \phi$ in a time $\tau$, the duration of the $\overline{\dot\phi^2}$ fluctuation in step 1. Thus
we require
\begin{equation}
|\overline{\dot\phi}| \geq \frac{|\Delta \phi|}{\tau} \approx \frac{a}{\tau} \,.
\label{eq:del-phi-min}
\end{equation}
In writing Eq.~\eqref{eq:Pstep1}, we assumed that $\ell \agt \tau$ and obtained a result which does not explicitly depend upon $\tau$. This means
that a typical $\overline{\dot\phi^2}$ fluctuation may be assumed to  last for a time which is of the order of $\ell$. For an estimate, we set $\tau \approx \ell$ in Eq.~\eqref{eq:del-phi-min}
and write
\begin{equation}
P_2(\overline{\dot\phi}) \approx  \exp( -  \ell^2 \, a^2)\,.
\label{eq:Pstep2b}
\end{equation}

Now we need the probability of the $\overline{\dot\phi^2}$ fluctuation and the $\overline{\dot\phi}$ fluctuation together, which depends upon whether the two fluctuations
are correlated. In lowest order, they are not, as $\langle \overline{\dot\phi}\; \overline{\dot\phi^2} \rangle =0$. However, there is a possibility of higher order correlations.
Here we assume that the two fluctuations are uncorrelated, so the probability of the two-step process depicted in  Fig.~\ref{fig:2-step} is
\begin{equation}
P_{12} = P_1(\overline{\dot\phi^2}) \; P_2(\overline{\dot\phi})\,.
\label{eq:P12}
\end{equation}
The decay rate per unit volume will again be taken to be
\begin{equation}
\Gamma_2 = \frac{P_{12}}{\tau \, \ell^3}\,
\label{eq:rate2}
\end{equation}
as in Eq.~\eqref{eq:rate1}.
Note that the anti-correlations discussed in Sect.~\ref{sec:anticorr} occur for the same operator measured at different times, and are not directly relevant here.

Now we wish to make some estimates of $P_{12}$. Consider the case $\alpha = 1/2$ and assume the thin wall approximation, so $\epsilon \ll 1$. We also assume that the spatial
averaging scale is of the order of  the initial bubble radius given in Eq.~\eqref{eq:tw-radius}, so
\begin{equation}
\ell \approx \frac{2}{ \epsilon \, a \,\sqrt{\lambda}} \,.
\end{equation}
We may combine this with Eq.~\eqref{eq:tw-parameters} to write
\begin{equation}
P_1(\overline{\dot\phi^2}) \approx  \exp\left( - \frac{2}{\epsilon^2 \,\sqrt{\lambda}} \right)\,.
\end{equation}
Similarly, we find
\begin{equation}
P_2(\overline{\dot\phi}) \approx  \exp\left( - \frac{4}{ \,\epsilon^2 \,\lambda} \right)\,.
\end{equation}
Note that here $P_2(\overline{\dot\phi})$ is much larger than the probability of the $\overline{\dot\phi}$ fluctuations considered in Sec.~II, as the
magnitude of the  $\overline{\dot\phi}$ fluctuation considered here is much smaller than that needed to lift a region over the barrier.

Now we have
\begin{equation}
P_{12} = {\rm e}^{-C}
\end{equation}
where
\begin{equation}
C \approx \frac{2}{ \,\epsilon^2 \,\lambda}\; \left(2 +\sqrt{\lambda} \right)\,.
\end{equation}
Comparison with Eq.~\eqref{eq:Btw} reveals
\begin{equation}
\frac{C}{B_{tw}} = \frac{3\, \epsilon}{4 \pi^2}\, (2 +\sqrt{\lambda})\,,
\end{equation}
$C < B$ in the thin wall approximation ($\epsilon \ll 1$), unless $\lambda$ is very large.
 Then the combined effects
of $\overline{\dot\phi^2}$ and $\overline{\dot\phi}$ fluctuations might dominate quantum tunneling.
This conclusion depends upon the assumption that the $\overline{\dot\phi^2}$ and $\overline{\dot\phi}$ fluctuations are
uncorrelated, or at least not strongly anti-correlated. This assumption needs further investigation.

Under this assumption, we may write the ratio of the decay rate due to  tunneling to that due to fluctuations as
\begin{equation}
\frac{\Gamma_I}{\Gamma_2} \approx  {\rm e}^{C -B + 5.78 - 10\, \ln \epsilon-2\,\ln\lambda}\,.
\label{eq:rate-ratio2}
\end{equation}
In the limit of small $\epsilon$ for fixed $\lambda$, this ratio approaches ${\rm e}^{-B} \ll 1$, so fluctuations give the dominant contribution to the decay rate.

\subsection{Measurement of $\overline{\dot\phi^2}$    }
\label{sec:measure}

In this subsection, we will describe a thought experiment by which an averaged quadratic operator such as $\overline{\dot\phi^2}$  could
be measured in a compact region of spacetime. Consider the Raychaudhuri equation for the expansion $\theta$ of a bundle of timelike
geodesics:
\begin{equation}
\frac{d\theta}{d \tau} = -R_{\mu\nu}\, u^\mu u^\nu\,,
\end{equation}
where $R_{\mu\nu}$ is the Ricci tensor, $u^\mu$ is the four-velocity of a particle on a geodesic, and $\tau$ is its proper time. Here
we have assumed that terms involving $\theta^2$ or the squares of the shear or vorticity may be neglected. If we measure the
change in the expansion of the bundle, $\Delta \theta$, averaged over finite intervals of time and space, then we have measured certain components
an averaged Ricci tensor,  $\overline{R_{\mu\nu}}$, averaged with compactly supported sampling functions $f(t)$ and $g({\bf x})$
defined by the details of the geodesic bundles. By varying the four-velocity, $u^\mu$, we can potentially obtain all of the diagonal
components of $\overline{R_{\mu\nu}}$. Next we may infer the averaged components of the stress tensor from the Einstein
equation in the form
\begin{equation}
\overline{T_{\mu\nu}} = \frac{1}{8\pi G} (\overline{R_{\mu\nu}} -\frac{1}{2} \, g_{\mu\nu}\,  \overline{R})\,,
\label{eq:Einstein}
\end{equation}
where $G$ is Newton's constant and $R = {R^\mu}_\mu$.

Now we assume that the source of the gravitational field is the self-coupled scalar field with Lagrangian density given in Eq.~\eqref{eq:Lagrange},
for which the stress tensor is
\begin{equation}
T_{\mu\nu} = \partial_\mu \phi \partial_\nu \phi -\frac{1}{2} g_{\mu\nu}\, \partial^\rho \phi \partial_\rho \phi  - g_{\mu\nu}\, U(\phi)\,.
\end{equation}
Further assume that the gravitational field is weak, and that in the above expression we may take the metric to have the Minkowski form
$g_{\mu\nu} \approx \eta_{\mu\nu} = diag(-1,1,1,1)$. If we form a particular combination of the components of $T_{\mu\nu}$, the potential
$U(\phi)$ cancels, and we have
 \begin{equation}
3\, T_{tt} + T_{xx} +T_{yy} +T_{zz} = 3\, {\dot\phi}^2 +  |\mathbf{\nabla} \phi|^2 \,.
\end{equation}
We expect ${\dot\phi}^2$ and $|\mathbf{\nabla} \phi|^2$ to be of the same order of magnitude, so measurements of $\overline{R_{\mu\nu}}$,
and hence of $\overline{T_{\mu\nu}}$, allow us to obtain an estimate of $\overline{\dot{\phi}^2}$ .

\subsection{What determines $\alpha$?}
\label{sec:alpha}

We have seen that the probability distributions for the fluctuations of quadratic operators, such as $\overline{\dot\phi^2}$ , are very sensitive to
the parameter $\alpha$ defined in Eq.~\eqref{eq:asyFourier}. This parameter determines the rate of decrease of $\hat f(\omega)$, the Fourier
transform of the temporal sampling function $f(t)$, and is also linked to the switch-on  and switch-off behavior of  $f(t)$. In the measurement
of $\overline{\dot\phi^2}$  described in the previous subsection, $\alpha$  will be determined by the details of the bundles of test particles
used. The more rapidly these bundles begin and end, the smaller will be  $\alpha$, and hence the larger the likely value of $\overline{\dot\phi^2}$
obtained in the measurement. This seems to imply that if we measure $\overline{\dot\phi^2}$  in the false vacuum state with bundles with small
$\alpha$, the probability of immediate decay is much greater than if we had use a larger value of $\alpha$. This issue needs further study, as it is
not immediately clear why a purely gravitational measurement should perturb the scalar field theory so much.

Another open question is what effect will fluctuations have upon the false vacuum in the apparent absence of a measurement of the form described
above. It is possible that the dynamics of coupling of the quantum field fluctuations with the classical background scalar field can determine a
specific value of $\alpha$, but how this might happen is unclear.

\subsection{Comparison of the Effects of Scalar and Electromagnetic Field Fluctuations}
\label{sec:comparison}

Recall that quantum electric field fluctuations have a small effect, of the order of $1\, \%$, on the rate of quantum tunneling of electrons through a
potential barrier~ \cite{ FZ99,HF15}. In contrast, we found in Sec.~\ref{sec:effects} that the effects of $\overline{\dot\phi}$ fluctuations can be
comparable to the rate of false vacuum decay as calculated in the instanton approximation, essentially a relative contribution of $O(1)$.
We can understand this difference as arising from the weakness of the electromagnetic interaction. The electric field fluctuation effect is
a one-loop correction to the tunneling rate, and is suppressed by a factor of the fine structure constant.

Similarly, the contribution of radiation pressure fluctuations to charged particle tunneling is very sensitive to the switching parameter $\alpha$,
and will be small compared to the WKB contribution unless $\alpha \alt 1/3$~\cite{HF16}.  By contrast, we argued above that $\overline{\dot\phi^2}$
fluctuations can give an $O(1)$ contribution to false vacuum decay.

As discussed in Sect.~\ref{sec:QM}, the instanton approximation seems to give a good
description of the Schwinger effect in both   $1+1$ and $3+1$ dimensions. However, the possible role of vacuum radiation pressure fluctuations
in the latter case merits further study. The Schwinger effect may be viewed as charged particle tunneling, so one expects the contributions
of both quantum electric field fluctuations and of radiation pressure fluctuations to be small. Thus one cannot use the agreement of instanton
and Bogolubov coefficient methods in the Schwinger effect to infer that $\overline{\dot\phi}$ or $\overline{\dot\phi^2}$ fluctuations will give a small
contribution to false vacuum decay.

\section{Summary and Conclusions}
\label{sec:final}

In this paper, we have discussed the effects of linear and quadratic quantum field fluctuations on the decay rate of a false vacuum
of a self-coupled scalar field. This rate is usually computed in an instanton approximation, where the solution of lowest Euclidean
action is assumed to dominate a path integral. We first consider the effects of the vacuum fluctuations of a linear field, $\dot{\phi}$,
averaged over finite intervals of both space and time. We have argued that this averaging describes a physical process or measurement
which necessarily begins and ends at finite times, and occurs in  compact regions of space and time. Hence the averaging should be
described by infinitely differentiable, but compactly supported and hence non-analytic functions of space and time. A quantum
$\dot{\phi}$ fluctuation in a finite region has an effect similar to a classical initial field velocity, and if its magnitude is large enough, can cause
a finite region to fly over a potential barrier, in a manner similar to the motion of a classical particle.

We find that quantum $\dot{\phi}$ fluctuations can cause false vacuum decay at a rate which can be comparable to the rate of quantum
tunneling, as described in the instanton approximation. This is consistent with the conclusions in
Refs~\cite{Linde92,CV99,CRV02,CRV01,ACRV03,CV06,BJPPW19,HY19,BDV19,Wang19},
although these authors offer differing conclusions as to whether linear quantum field fluctuations are an alternative formalism for
describing quantum tunneling, or represents a distinct physical process. We adopt the latter viewpoint.
Evidence that  $\dot{\phi}$
fluctuations are a separate decay channel from tunneling arises in the wide variation in decay rates, as opposed to the logarithm of the rates
found in Sect.~\ref{sec:num}. Further evidence comes from the
 dependence of the variance in Eqs.~\eqref{eq:sigma2} and
\eqref{eq:eta}  upon the sampling functions. Our view is that these functions should be determined by the physical details of the
system, here perhaps the dynamics of the formation of the bubble of true vacuum.

This dependence is even more pronounced in the case of the effects of quadratic quantum field fluctuations, such as  $\overline{\dot\phi^2}$,
upon the decay rate. Here we found an effect which can be significantly more likely than  large $\overline{\dot{\phi}}$ fluctuations.
This arises because the probability distribution, $P(\overline{\dot\phi^2})$, falls more slowly than an exponential function  when compactly supported
averaging functions are used, and provides evidence that both  $\overline{\dot{\phi}}$ and  $\overline{\dot\phi^2}$ fluctuations provide  different
decay processes than quantum tunneling. However,
our analysis in Sec.~\ref{sec:effects2} relies upon an assumption, Eq.~\eqref{eq:P12}, concerning the correlation of linear and quadratic
operator fluctuations which need to be examined further. If this assumption is correct, then at least for
false vacuum decay in $3+1$ dimensions, quadratic operator fluctuations may be the dominant decay mechanism. The role of such fluctuations
in other contexts remains to be explored  in more detail.

\begin{acknowledgments}
We would like to thank Mark Hertzberg, Ken Olum, Alex Vilenkin, Shao-Jiang Wang, and Masaki Yamada for helpful discussions. We
also thank Ali Masoumi for providing the software package described in Ref.~\cite{MOS17}, and assistance in its use.
This work was supported in part  by the National Science Foundation under Grant PHY-1912545.
\end{acknowledgments}


\begin{thebibliography}{20}

\bibitem{FZ99} V.V  Flambaum and V.G. Zelevinsky, Radiation corrections increase tunneling probability,
Phys. Rev. Lett. {\bf 83}, 3108 (1999), arXiv:nucl-th/9812076.

 \bibitem{HF15} H. Huang and L.  H. Ford, Quantum electric field fluctuations and potential scattering,
 Phys. Rev. D {\bf 91}, 125005 (2015), arXiv:1503.02962.

 \bibitem{HF16} H. Huang and L.  H. Ford, Vacuum radiation pressure fluctuations and barrier penetration,
 Phys. Rev. D {\bf 96}, 016003 (2017), arXiv:1610.01252.

 \bibitem{F21} L.  H. Ford,  Vacuum Radiation Pressure Fluctuations on Atoms,  Phys. Rev. A {\bf104}, 012208 (2021),
 arXiv:2104.03212.

 \bibitem{C77} S. Coleman, Fate of the false vacuum: Semiclassical theory, Phys. Rev. D {\bf 15}, 2929 (1977).

  \bibitem{Linde92} A. Linde,  Stochastic approach to tunneling and baby universe formation,  Nucl. Phys. {\bf B372}, 421 (1992), hep-th/910037.

 \bibitem{CV99} E. Calzetta,  and E. Verdaguer, Noise induced transitions in semiclassical cosmology,
 Phys. Rev. D {\bf 59}, 083513 (1999), arXiv:gr-qc/9807024.

 \bibitem{CRV02} E. Calzetta, A. Roura,  and E. Verdaguer, Dissipation, noise and vacuum decay in quantum field theory,
 Phys. Rev. Lett. {\bf 88}, 010403 (2001), arXiv:hep-ph/0101052.

 \bibitem{CRV01} E. Calzetta, A. Roura,  and E. Verdaguer, Vacuum decay in quantum field theory,
 Phys. Rev. D {\bf 64}, 105008 (2001), arXiv:hep-ph/0106091.

 \bibitem{ACRV03} D. Arteaga, E. Calzetta, A. Roura,  and E. Verdaguer, Activation-like processes at zero temperature,
Int. J. Theor. Phys. {\bf 42}, 1257 (2003), arXiv:quant-ph/0303075.

 \bibitem{CV06}  E. Calzetta,  and E. Verdaguer, Real time approach to tunneling in open quantum systems: decoherence and
anomalous diffusion, J.Phys.A {\bf 39} 9503 (2006), arXiv:quant-ph/0603047.

 \bibitem{seminar18} H. Huang,  False vacuum decay and quantum fluctuations,  seminar at Tufts University,
 April 19, 2018.

 \bibitem{BJPPW19} J. Braden, M. C. Johnson, H. V. Peiris, A. Pontzen, and S. Weinfurtner,  A new semiclassical
picture of vacuum decay,  Phys. Rev. Lett. {\bf 123}, 031601 (2019), arXiv:1806.06069.

 \bibitem{HY19} M. P. Hertzberg and M. Yamada, Vacuum decay in real time and imaginary time formalisms,
 Phys. Rev. D {\bf 100}, 016011 (2019), arXiv:1904.08565.

 \bibitem{BDV19} J. J. Blanco-Pillado, H. Deng, and A. Vilenkin, Flyover vacuum decay,  JCAP 12  (2019) 001, arXiv:1906.09657.

 \bibitem{Wang19} S.-J. Wang, Occurrence of semiclassical vacuum decay,  Phys. Rev. D {\bf  100}, 096019 (2019),
 arXiv:1909.11196.

\bibitem{FFR12}  C. J. Fewster, L. H. Ford, and T. A. Roman, Probability distributions for quantum stress tensors in four dimensions,
Phys. Rev. D {\bf 85}, 125038 (2012),  arXiv:1204.3570.

\bibitem{FF15}  C. J. Fewster and L. H. Ford,  Probability distributions for quantum stress tensors measured in a finite time interval,
Phys. Rev. D {\bf 92}, 105008 (2015),  arXiv:1508.02359.

\bibitem{SFF18}  E. D. Schiappacsse, C. J. Fewster and L. H. Ford,  Vacuum quantum stress tensor fluctuations:
A diagonalization approach, Phys. Rev. D {\bf 97}, 025013 (2018);, arXiv:1711.09477.

\bibitem{FF19}  C. J. Fewster and L. H. Ford,  Probability distributions for space and time averaged quantum stress tensors,
Phys. Rev. D {\bf 101}, 025006 (2020), arXiv:1909,07285.

 \bibitem{WFS21} P. Wu, L. H. Ford, E. D. Schiappacasse, Space and Time Averaged Quantum Stress Tensor Fluctuations,
   Phys. Rev. D {\bf 103}, 125014 (2021),   arXiv:2104.04446.

  \bibitem{C85}  S. Coleman,  {\it Aspects of Symmetry} (Cambridge University Press,
  Cambridge, 1985),  Chap. 7, The Uses of Instantons.

 \bibitem{Dunne2000} G.V. Dunne and K. Rao, Lam{\'e}  Instantons, JHEP 01, 019 (2000).

 \bibitem{M-K01} H.J.W. M{\"u}ller-Kirsten, J-z. Zhang, and Y. Zhang,  Once again: instanton method vs.WKB,  JHEP 11, 011 (2001)

 \bibitem{Benderskii08} V.A. Benderskii, E.V. Vetoshkin, and E.I. Kats, Accuracy of semiclassics: Comparative analysis of WKB and instanton approaches,
 HAIT Journal of Science and Engineering A {\bf 5} 71-92 (2008),  arXiv:cond-mat/0506027.

 \bibitem{Schwinger51} J. Schwinger, Gauge invariance and vacuum polarization, Phys. Rev. {\bf 82}, 664-679 (1951).

 \bibitem{Garriga94a} J. Garriga, Nucleation rates in flat and curved space,  Phys. Rev. D {\bf 49},
 6327-6342 (1994).

\bibitem{KP02}  S.P. Kim and D.N. Page, Schwinger pair production via instantons in strong electric fields,  Phys. Rev. D {\bf 65},
105002 (2002).

\bibitem{Garriga94b} J. Garriga, Pair production by an electric field in (1+1)-dimensional de Sitter space,  Phys. Rev. D {\bf 49},
 6343-6346 (1994).

 \bibitem{Grib94} A.A. Grib, S.G. Mamayev, and V.M. Mostepanenko, {\it Vacuum Quantum Effects in Strong Fields}
  (Frienmann Laboratory Publishing, St. Petersberg, 1994) Chap. 4.

 \bibitem{CC77} C. G. Callan and S. Coleman, Fate of the false vacuum II: First quantum corrections,
 Phys. Rev. D {\bf 16}, 1762 (1977).

\bibitem{MOS17}   A. Masoumi, K. Olum, and B. Shlaer, Efficient numerical solution to vacuum decay with many fields,
JCAP {\bf 01} 051 (2017), arXiv:1610.06594.

 \bibitem{GM15} B. Garbrecht and Peter Millington, Greens function method for handling radiative effects on false vacuum decay,
 Phys. Rev. D {\bf 91}, 105021 (2015), arXiv:1501.07466.

 \bibitem{FR05} L. H. Ford, and T. A. Roman, Minkowski Vacuum Stress Tensor Fluctuations, Phys. Rev. D {\bf 72},
 105010 (2005),  arXiv:qr-qc/0506026.

 \bibitem{PF11} V. Parkinson and L. H. Ford, A Model for Non-Cancellation of Quantum Electric Field Fluctuations,
 Phys. Rev. A  {\bf 84}, 062102 (2011), arXiv:1106.6334.

\bibitem{FFR10} C.J.~Fewster, L.H.~Ford and T.A.~Roman, ``Probability distributions of smeared
 quantum stress tensors,'' Phys. Rev. D \textbf{81}, 121901 (2010), arXiv:1004.0179 [quant-ph].

 \bibitem{KM84} F. R. Klinkhamer and N.S. Manton, A saddle-point solution in the Weinberg-Salam theory,
 Phys. Rev. D {\bf 30}, 2212 (1984).

\bibitem{AM87-88} P. Arnold and L.McLerran, Sphalerons, small fluctuations, and baryon-number violation in electroweak theory,
Phys. Rev. D {\bf 36}, 581 (1987);
The sphaleron strikes back: A response to objections to the sphaleron approximation, Phys. Rev. D {\bf 37}, 1020 (1988).


 \end{thebibliography}
\end{document}